\documentclass[english,11pt,oneside]{article}
\pdfoutput=1
\usepackage{amsmath}
\usepackage{amsfonts}
\usepackage{amssymb}
\usepackage{graphicx, rotating}
\usepackage{epstopdf}
\usepackage{epsfig}
\usepackage{latexsym}
\usepackage{multirow}
\usepackage{color}
\usepackage{slashed}
\usepackage{pifont}
\usepackage[top=2.8 cm, bottom=2.8 cm, left=3 cm, right=3 cm]{geometry}
\usepackage[compat=1.0.0]{tikz-feynman}

\newcommand{\ba}{\begin{eqnarray}}
\newcommand{\ea}{\end{eqnarray}}
\newcommand{\no}{\nonumber}

\usepackage{hyperref}
\hypersetup{
 colorlinks = true,
 linkcolor  = red,
 citecolor={blue}
}

\usepackage{bbold}

\begin{document}

\title{
Axion Quality Straight from the GUT
}
\author{Luca Vecchi~\footnote{luca.vecchi@pd.infn.it}\\
{\small\emph{Istituto Nazionale di Fisica Nucleare (INFN), Sezione di Padova, Italy}}}
\date{}
\maketitle

\begin{abstract}

Composite axion scenarios offer a robust field theoretic justification for the existence of a Peccei-Quinn symmetry of high quality. We present a new class of realizations that are naturally embedded in Grand-Unified Theories, retain asymptotic freedom for all gauge groups, and protect the axion symmetry up to operators of dimension 12. Our setup leads to a number of distinctive signatures at low energies. First, additional composite scalars are predicted; some of these are viable dark matter candidates for values of the axion decay constant that are too low for the QCD axion abundance to be relevant. Second, an approximate unification of the Standard Model gauge couplings takes place at the axion scale, while leaving the actual quark-lepton unification at much higher energies as usual. This suggests the existence of GUT relics with Standard Model gauge quantum numbers at potentially accessible scales.

\end{abstract}

\newpage

{
	\hypersetup{linkcolor=black}
	\tableofcontents
}

\section{Introduction}
\label{sec:intro}

The non-observation of the electric dipole moment of the neutron sets an upper bound on the QCD topological angle: 
\ba\label{boundTheta}
|\bar\theta|\lesssim10^{-10}
\ea
The smallness of $|\bar\theta|$ can be explained by postulating non-generic properties of the UV completion of the Standard Model (SM). The identification of such properties is a challenging theoretical and experimental task, called Strong CP Problem. Among the various solutions of the Strong CP Problem are UV completions that invoke a global $U(1)_{\rm PQ}$ anomalous under QCD. 

Within the new physics sector the symmetry $U(1)_{\rm PQ}$ may be linearly realized, explicitly broken, or spontaneously broken. When the $U(1)_{\rm PQ}$ is linearly realized, $\bar\theta$ can be rotated away becoming unphysical. By anomaly matching, once the massive new physics states are integrated out the low energy spectrum necessarily includes (composite or fundamental) massless colored fermions. Unfortunately, experiments firmly preclude the possibility of exotic massless quarks, and also the massless up-quark hypothesis~\cite{masslessU}  appears to be rather strongly rejected (see, e.g., \cite{Aoki:2013ldr} and more recently \cite{Alexandrou:2020bkd}). If instead the $U(1)_{\rm PQ}$ is explicitly broken by the new physics, the full theory has a CP-odd angle $\bar\theta_{\rm full}=\bar\theta-\bar\theta'$ constructed as a linear combination of the QCD topological angle and some other anomalous parameter $\bar\theta'$ of the new physics. Postulating a softly-broken $Z_2$ mirror symmetry such that $\bar\theta'=\bar\theta$ makes the overall phase $\bar\theta_{\rm full}$ vanish.~\cite{Hook:2014cda} Because this is the quantity we actually measure below the mirror sector scale, the strong CP problem is solved. To be compatible with collider searches, explicit realizations of this idea must ensure the scale of soft $Z_2$ breaking lies within a certain range. 

In the most popular $U(1)_{\rm PQ}$-based approach to the Strong CP Problem, the anomalous Peccei-Quinn symmetry is spontaneously broken and an approximate Nambu-Goldston boson, the QCD axion $a$, is predicted.~\cite{Peccei:1977hh} An anomalous rotation of the colored fermions does not remove the topological angle; rather, it shifts the axion in such a way as to keep the dynamical field $\bar a/f_a=\bar\theta+a/f_a$ invariant. Remarkably, the vacuum energy $V_{\rm QCD}(\bar a)$ of QCD gets minimized at $\langle\bar a\rangle=0$. This means that the value of the observable topological angle gets automatically relaxed to zero, $\bar\theta_{\rm obs}\equiv\langle\bar a\rangle=0$, provided the axion potential is entirely dominated by non-perturbative QCD effects.

The QCD axion has been sought after for many years. No evidence has ever been found and the experimental searches are interpreted as constraints on the parameter space. Because all axion couplings are controlled by the scale $f_a$ of spontaneous $U(1)_{\rm PQ}$ breaking, data translate into lower bounds on this quantity. For the invisible axion models considered here, the most stringent constraints read $f_a\gtrsim10^8$ GeV. Such large scales make the axion potential extremely sensitivity to tiny effects from unknown physics at the shortest imaginable distances.\cite{Holman:1992us} 

Being an anomalous symmetry, $U(1)_{\rm PQ}$ cannot be exact in the far UV. At best, we may expect it to be broken solely by quantum gravity effects parametrized at low energies via Planck-scale suppressed interactions. These induce corrections $\Delta V$ to the axion potential that are generically not aligned with $V_{\rm QCD}(\bar a)$, the origin of field space is destabilized and a non-vanishing observable $\bar\theta_{\rm obs}\equiv\langle\bar a\rangle/f_a$ is re-introduced. The full potential is now the sum of a term due entirely to QCD and one from tiny effects at the Planck scale:
\ba\label{eq1V}
V(\bar a)=V_{\rm QCD}(\bar a)+\Delta V(\bar a)
\ea
At leading order in $\Delta V$ we find
\ba\label{DeltaTheta}
\bar\theta_{\rm obs} =-\left.\frac{\Delta V'}{f_aV_{\rm QCD}''}\right|_{\langle\bar a\rangle=0}.
\ea
The question of why $|\bar\theta_{\rm obs}|$ is so small becomes, in axion models, the question of why the $U(1)_{\rm PQ}$ is so accurate despite it being only an approximate symmetry of the UV dynamics. 

This ``$U(1)_{\rm PQ}$ quality problem" is so significant because the QCD potential is controlled by physics at the GeV scale, $V_{\rm QCD}\sim m_\pi^2f_\pi^2$, whereas $\Delta V$ is a function of the large $f_a$ and the Planck scale $f_{\rm Pl}=2.4\times10^{18}$ GeV. Using naive dimensional analysis we expect $\Delta V\sim c\,16\pi^2f_a^4(f_a/f_{\rm Pl})^{d-4}$, with $c$ some coefficient and $d$ the dimension of the associated $U(1)_{\rm PQ}$-violating operator. For a generic, maximally strong UV dynamics $c\sim1$. As long as $f_a\ll f_{\rm Pl}$ one can hope to suppress the destabilizing effect by making sure the operator is sufficiently irrelevant. Plugging these estimates in \eqref{DeltaTheta}, the experimental bound \eqref{boundTheta} translates into $d\gtrsim9$ for $f_a\sim10^8$ GeV. A scale of order, say, $f_a\sim10^{11}$ GeV would instead require $d\gtrsim12$. As $f_a$ approaches the Planck scale the only way to prevent $|\bar\theta_{\rm obs}|$ from exceeding the experimental limit is to have $|c|\lesssim10^{-10}$. Unfortunately, to a low energy observer such a condition is no different than the experimental input \eqref{boundTheta} we started with. The necessary non-generic UV conditions to ensure $|c|\ll1$ may be somehow realized in the correct theory of quantum gravity, for example in String Theory. However, in this paper we take the point of view of the low energy observer and look for explanations within the realm of quantum field theory.

We see that robust realizations of the QCD axion within a field theory approach require two fundamental ingredients: an axion decay constant parametrically smaller than the Planck scale and an accurate $U(1)_{\rm PQ}$. Composite axion models can feature both ingredients. In such a framework the axion arises as a Nambu-Goldstone boson (NGB) of some exotic strong dynamics that undergoes chiral symmetry breaking at a scale $\sim f_a$.~\cite{Kim:1984pt} As long as the axion constituents are chiral under the new gauge interactions, a number of accidental chiral symmetries, among which $U(1)_{\rm PQ}$, may be very accurate. Furthermore, a value of $f_a$ that is simultaneously parametrically larger than the weak scale and lower than the Planck scale is completely natural in composite axion models, where $f_a$ gets generated via dimensional transmutation. In non-supersymmetric models with fundamental scalars, on the other hand, the value of $f_a$ results from a fine-tuning that is much more severe than just setting $|\bar\theta|<10^{-10}$ by hand in a given UV completion of the SM.

The first composite axion model with chiral constituents was proposed in \cite{Randall:1992ut}. In those types of models the fermion constituents are charged under three non-abelian gauge symmetries and to have a sufficiently accurate $U(1)_{\rm PQ}$ some of the gauge groups must have rank bigger than a certain lower bound. As a consequence there is a structural tendency to developing low energy Landau poles.~\cite{Dobrescu:1996jp} One can evade this problem by replacing one or more of these non-abelian factors with abelian groups \cite{Fukuda:2017ylt} (one should also worry about cubic anomalies in this case). The price to pay is however a loss of asymptotic freedom. Alternatively, one can construct asymptotically-free models using a ``moose" gauge structure \cite{Redi:2016esr} that resembles axion scenarios in a local extra dimension~\cite{Cheng:2001ys}.

The purpose of the present paper is proposing an alternative class of asymptotically-free composite axion models with a natural embedding in some Grand-Unified Theory (GUT) and a $U(1)_{\rm PQ}$ protected up to operators of dimension $d=12$.

\section{Warming up with $SU(N_c)\times SU(5)^2$}
\label{sec:su5}

We begin our discussion with a very simple model. Unfortunately its protection of the $U(1)_{\rm PQ}$ is at most barely sufficient. Yet, this toy model allows us to illustrate some of the key features that characterize the more efficient scenarios discussed in Section \ref{sec:so10}.

Our toy-model is based on an $SU(N_c)\times SU(5)_1\times SU(5)_2$ gauge symmetry. The axion constituents are shown in Table \ref{modSU5}. They are all Weyl fermions, chiral under $SU(5)^2$. The $SU(N_c)$ is the dynamics that becomes strong and generates a composite axion at scales of order $f$. The $SU(5)_1$ is to be identified with the GUT, with the SM fermions belonging to three families in the ${\bf 10}\oplus\overline{\bf 5}\in SU(5)_1$ and being singlets of $SU(N_c)\times SU(5)_2$. The $SU(5)_1$ GUT is not instrumental to the realization of our composite axion scenarios. Yet, being one of our motivations we find it useful to entertain the possibility of grand-unification. In this case scalar fields at the GUT scale should be added to obtain a realistic model. The axion physics in our models is not going to be affected by them provided they do not carry $SU(N_c)\times SU(5)_2$ charges, which is what we assume. No other fundamental scalar is introduced.~\footnote{We have nothing new to say about the generation and the stabilization of the hierarchy between the weak and the GUT scales. The hierarchy problem is assumed to be addressed in some way. On our part, the axion models proposed here do not introduce new fine-tunings; in particular, the new mass scale $f_a$ is dynamically generated and can be naturally smaller than the Planck mass.}

\begin{table}[!t]
	\begin{center}
		\begin{tabular}{c||ccc||ccc}
			& $SU(N_c)$ & $SU(5)_1$ & $SU(5)_2$ & $U(1)_{\rm B}$ & $U(1)_{\rm V}$ & $U(1)_{\rm PQ}$ \\
			\hline\hline
			$\psi$ & ${\bf N_c}$ & ${\bf 10}$ & ${\bf 1}$ & $1$ & $1$ & $1$ \\
			${\widetilde\psi}$ & ${\bf N_c}$ & $\overline{\bf 5}$ & ${\bf 1}$ & $1$ & $-2$ & $-2$ \\
			$\psi^c$ & $\overline{\bf N_c}$ & ${\bf 1}$ & $\overline{\bf 10}$ & $-1$ & $-1$ & $1$ \\
			${\widetilde\psi}^c$ & $\overline{\bf N_c}$ & ${\bf 1}$ & ${\bf 5}$ & $-1$ & $2$ & $-2$
		\end{tabular}
		\caption{Axion constituents, and their accidental $U(1)$ symmetries, in our toy-model.}\label{modSU5}
	\end{center}
\end{table}

At scales of order $f_{\rm GUT}$ the GUT is broken into the SM as usual:
\ba\label{GUTthreshold1}
SU(5)_{1}&\to& G_{\rm SM,1}\\\no
&\equiv& SU(3)_{G,1}\times SU(2)_{W,1}\times U(1)_{Y,1}
\ea
At $f\ll f_{\rm GUT}$, when the strong dynamics of the axion becomes important, it is appropriate to decompose $\psi,\widetilde\psi$ in irreducible $G_{\rm SM,1}$ representations: $\psi=\psi_{q}\oplus\psi_u\oplus\psi_e$, $\widetilde\psi=\psi_d\oplus\psi_\ell$, where $\psi_{q,u,e,d,\ell}$ are in the fundamental of $SU(N_c)$, singlets under $SU(5)_{2}$ and transform under $G_{\rm SM,1}$ as the SM quarks and leptons $q,u,e,d,\ell$.

Let us next discuss under which conditions our gauge interactions remain asymptotically free. The requirement of asymptotic freedom for $SU(5)_{2}$ is satisfied for any $N_c\leq13$. This is a very mild constraint. A much more significant upper bound on $N_c$ is found demanding that $SU(5)_{1}$ remains UV-free. This bound is model-dependent because it depends on the additional fields necessary to explain \eqref{GUTthreshold1} and the SM Yukawa couplings. For definiteness we consider a minimal (non-supersymmetric) scenario in which the scalar responsible for breaking $SU(5)_{1}$ into the SM is in the representation ${\bf 24}$. The observed pattern of Yukawa couplings for the SM fermions may be reproduced with a Higgs in the ${\bf 5}$. Rather than introducing additional fundamental fields, which would quickly take our model away from asymptotic freedom, the necessary quark-lepton splitting is achieved via effective dimension-5 operators with two SM fermions and two powers of the scalars  $\phi_H\sim{\bf 24}\oplus{\bf5}$. These interactions are large enough provided 
\ba\label{GUTscale}
f_{\rm GUT}\gtrsim10^{16}~{\rm GeV}.
\ea
With this minimal GUT-scale field content the upper bound reads $N_c\leq9$. The key assumption that $SU(N_c)$ confines and breaks its chiral symmetries provides instead a lower bound on $N_c$. Ignoring the weak gauging of $SU(5)^2$, the $SU(N_c)$ dynamics behaves as if it was a QCD-like theory with $N_c$ colors and $N_f=15$ massless flavors. According to a number of independent arguments (for an incomplete list see, e.g., \cite{Appelquist:1996dq}\cite{Appelquist:1999hr}\cite{Poppitz:2009uq}\cite{Ryttov:2017lkz}\cite{Gardi:1998ch}), confinement and dynamical chiral symmetry breaking occurs when $N_f/N_c\lesssim 3-4$. We will work in the regime $5\leq N_c\leq9$. 

Under our assumptions, at $f\sim10^8-10^{9}$ GeV the $SU(N_c)$ becomes strong and non-trivial vacuum condensates form:
\ba\label{condensates1}
\frac{1}{10}\langle \psi\psi^c\rangle=\frac{1}{5}\langle {\widetilde\psi}{\widetilde\psi}^c\rangle\sim 4\pi f^3\sqrt{\frac{3}{N_c}},
\ea
where we borrowed the factors of $4\pi$ from QCD and the powers of $N_c$ from large $N_c$ scaling.~\footnote{For QCD in the chiral limit we have $\langle \psi\psi^c\rangle\simeq(240~{\rm MeV})^3$ at $\mu\sim1$ GeV (no summation over flavor). This agrees with the naive dimensional analysis estimate $\langle \psi\psi^c\rangle\sim4\pi f_\pi^3$ up to a number close to 1.} The particle content of Table \ref{modSU5} features a number of accidental symmetries. We spare the reader of the details, which will instead be presented for the models of Section \ref{sec:so10}. The bottom line is that \eqref{condensates1} break the gauge symmetry $G_{\rm SM,1}\times SU(5)_2$ into the diagonal $G_{\rm SM}$, which is to be identified with the SM gauge symmetry we experimentally observe. In the process, many Nambu-Goldstone bosons (NGBs) are generated. Some are eaten via the Higgs mechanism, some acquire large masses squared $\sim g_{\rm SM}^2f^2/N_c$, and finally four SM-neutral NGBs remain massless at the renormalizable level. Among these, three are associated to chiral symmetries that appear below the GUT scale but are in fact explicitly broken by the weak gauging of $SU(5)^2$, which explains why these are not shown in Table \ref{modSU5}. As a consequence, these three NGBs acquire masses from dimension-6 interactions below the GUT scale (we will study them in Section \ref{sec:unstableALP}). The actual QCD axion $a$ is associated to the breaking of the anomalous $U(1)_{\rm PQ}$ of Table \ref{modSU5}.

The lowest dimensional operators that violate $U(1)_{\rm PQ}$ are formed by the product of $\psi^\dagger\bar\sigma^\mu\widetilde{\psi}$ and appropriate SM fermion currents, and have dimension $d=6$. Yet, such interactions are innocuous. Indeed, to generate an axion potential $\Delta V$ in our QCD-like models one must break {\emph{both}} $U(1)_{\rm PQ}\subset SU(N_f)_L$ as well as $U(1)_{\rm PQ}\subset SU(N_f)_R$. In the presence of explicit breaking of $SU(N_f)_L$ only, as it is the case for the dimension-6 interactions just mentioned, one could remove completely from the Lagrangian the non-derivative couplings of the axion performing a global $SU(N_f)_R$ rotation. This demonstrates that the above dimension-6 interactions are not harmful unless they are accompanied by operators that also violate the $U(1)_{\rm PQ}\subset SU(N_f)_R$ carried by $\psi^c,\widetilde\psi^c$. Fortunately, the couplings of the latter fields are more severely constrained by invariance under $SU(5)_2$ and $SU(N_c)$, as we now argue. 

The leading $SU(5)_2$-invariant combinations that can break $U(1)_{\rm PQ}\subset SU(N_f)_R$ are $\psi^c\widetilde\psi^c\widetilde\psi^c$ and $\psi^c\psi^c(\widetilde\psi^c)^\dagger$. The former requires three powers of $\psi,\widetilde\psi$ to make an $SU(N_c)$-singlet, leading us to operators of at least dimension $d=9$. The other option, i.e. $\psi^c\psi^c(\widetilde\psi^c)^\dagger$, may be combined with $\psi\psi(\widetilde\psi)^\dagger$, which again gives $d=9$ operators. Hence, we conclude that $\Delta V$ is first induced by operators of the type
\ba\label{SU5eff}
(\psi\psi^c)(\widetilde\psi\widetilde\psi^c)^2,~~~~~~(\psi\psi^c)^2(\widetilde\psi\widetilde\psi^c)^\dagger.
\ea
Strictly speaking, one may also build an $SU(N_c)$ singlet from $\psi^c\psi^c(\widetilde\psi^c)^\dagger$ by adding either $\psi$ or $\widetilde\psi$  (or even their conjugates if we are willing to pay the additional price of a gauge field strength), but the resulting operator would not be invariant under $G_{\rm SM, 1}$ unless at least an additional $SU(N_c)$-singlet, say a Higgs and/or an even number of SM fermions, are added. 
One can verify that the effect of these operators on the axion potential is negligible compared to the one coming directly from \eqref{SU5eff}.

Having established that the dominant contribution to $\Delta V$ arises from the set of operators in eq. \eqref{SU5eff}, we can finally estimate the expected size of the observed topological angle \eqref{DeltaTheta}. For later convenience we analyze the impact of a general operator of dimension $d$ of the type $(\Psi\Psi^c)^{d/3}$, with $\Psi,\Psi^c$ the axion constituents. We expect the operator to be suppressed by the ultimate cutoff, say the Planck scale $f_{\rm Pl}\simeq2.4\times10^{18}$. Naive dimensional analysis is used to estimate the Wilson coefficient. Denoting by $g_{\rm Pl}$ the typical coupling at the cutoff, e.g. the string coupling, the operator appears in the effective Lagrangian as
\ba\label{operGen}
{\cal L}_{\slashed{\rm PQ}}&\supset&\frac{c_{\rm Pl}}{g^{d/3-2}_{\rm Pl} f_{\rm Pl}^{d-4}}(\Psi\Psi^c)^{d/3}+\cdots,
\ea
where $c_{\rm Pl}$ is of order unity if we conservatively assume the operator is generated at tree-level by exotic states of mass $\sim g_{\rm Pl}f_{\rm Pl}$. For any $d>6$ the Wilson coefficient gets smaller as the coupling increases (we have fixed $f_{\rm Pl}$). Large-N counting suggests that the maximal allowed value satisfies $g_{\rm Pl}^2N_{\rm dof}\sim16\pi^2$, with $N_{\rm dof}\gg1$ the number of degrees of freedom. In our estimates we will saturate the bound with $N_{\rm dof}\sim N_fN_c$. Fortunately, our numerical results are modestly affected by the unknown parameter $g_{\rm Pl}$.

One should then run the Wilson coefficient down to energies of order the confinement scale, i.e. $c_{\rm Pl}\to c_{\rm IR}$. We may estimate this effect assuming (as in the large $N_c$ approximation) the operator anomalous dimension factorizes as $d/3$ times the anomalous dimension of the quark bilinear $\Psi\Psi^c$. Limiting our analysis to a domain $g^2_{\rm IR}N_c/(4\pi)<0.5-1$ in which a perturbative calculation is expected to be reliable, we observe an enhancement of order $c_{\rm IR}/c_{\rm Pl}\sim10-10^2$. The contribution to the axion potential due to \eqref{operGen} can thus be written as 
\ba
\Delta V=
\frac{c_{\rm IR}}{g_{\rm Pl}^{d/3-2}f_{\rm Pl}^{d-4}}n_{\cal O}\left(4\pi f^3\sqrt{\frac{3}{N_c}}\right)^{d/3}~e^{2iq_a a/f}+{\rm hc}~~~~
\ea
where $q_a$ is the $U(1)_{\rm PQ}$ charge of \eqref{operGen} and $n_{\cal O}\propto N_f^{d/3}$ is a numerical factor that depends on the number of condensates \eqref{condensates1} that appear in the operator \eqref{operGen}. For example, the second operator in \eqref{SU5eff} contributes with $n_{\cal O}=10^2\times5^1$ and $q_a=4/\sqrt{15}$. For generic complex coefficients $c_{\rm IR}$, uncorrelated with the bare $\bar\theta$, $\Delta V$ implies a contribution to the QCD topological angle (see eq. \eqref{DeltaTheta}) 
\ba\label{estimateThetaGen}
|\bar\theta_{\rm obs}|
&=&\frac{(m_u+m_d)^2}{m_um_d}\frac{(4\pi)^2f_a^4}{m_\pi^2f_\pi^2}\left(\frac{f_a}{f_{\rm Pl}}\right)^{d-4}\\\no
&\times&4n_{\cal O}q_a~|c_{\rm IR}|\left|\sin\left({\rm Arg}(c_{\rm IR})-\bar\theta\right)\right|~\left(\frac{4\pi}{g_{\rm Pl}}\right)^{d/3-2}({\cal A}_a^G)^{d-1}\left(\frac{3}{N_c}\right)^{d/6}.
\ea
Here we employed the leading order expression for the axion mass $m_a^2=V''_{\rm QCD}=[m_um_d/(m_u+m_d)^2]m_\pi^2f_\pi^2/f_a^2$, with $m_u/m_d\simeq0.48$, and used the definition $f={\cal A}_a^Gf_a$, where ${\cal A}_a^G$ is the QCD anomaly coefficient determining the axion coupling to gluons (see Section \ref{sec:axion} for details). Note that the powers of ${\cal A}_a^G=N_c/\sqrt{15}\propto N_c$ and $N_f$ (implicitly included in $n_{\cal O}$), as well as the RG effects, go in the direction of enhancing $|\bar\theta_{\rm obs}|$. Other than that, the estimate \eqref{estimateThetaGen} agrees very well with the one given in Section \ref{sec:intro}. 

Employing typical values of $n_{\cal O}$ and $q_a$ (such as those that have been mentioned above \eqref{estimateThetaGen}), the constraint \eqref{boundTheta} reads $f_a\lesssim10^{8}$ GeV, which falls too short to convince us this is a viable model. One can certainly hope that the actual value of the unknown coefficient $c_{\rm Pl}$ may be suppressed in a realistic theory of gravity. But this is not what we do here. Our point of view is the one of an agnostic low energy observer who would like to find an answer to a puzzling result, i.e. \eqref{boundTheta}, without having to resort to unjustified assumptions about the UV. In other words, we are interested in finding scenarios that work with coefficients $c_{\rm Pl}$ of order unity. Such realizations are presented in Section \ref{sec:so10}.

Before concluding this discussion we would like to stress that our toy model differs significantly from the one of Ref. \cite{Gavela:2018paw}, where the axion constituents are also in a chiral ${\bf 10}\oplus\overline{\bf 5}$ representation of some $SU(5)$. In that paper the gauge group $SU(5)$ is identified with the strong axion dynamics; the constituents are therefore chiral under the confining force but vector-like under the SM. Here the situation is completely reversed. Our axion constituents carry vector-like representations under the confining $SU(N_c)$ dynamics and are instead chiral under the weakly-gauged, grand-unified $SU(5)^2$. The dimensionality of the leading operators that violate $U(1)_{\rm PQ}$ is inherited from the chiral ${\bf 10}\oplus\overline{\bf 5}\in SU(5)$, and is thus the same here as well as \cite{Gavela:2018paw}, though.

\section{The $SU(N_c)\times SO(10)^4$ Model}
\label{sec:so10}

The previous model does not quite reach the desired level of protection of the axion $U(1)_{\rm PQ}$. The ultimate reason is that it is sufficient to have three powers of ${\bf 10}\oplus{\overline{\bf5}}\in SU(5)$ to build a complete gauge singlet under the chiral $SU(5)^2$. Our next models prevent this from happening by embedding the axion constituents in the spinorial representation of $SO(10)$. Apart from a more significant suppression of $\Delta V$, in many respects the new models are qualitatively similar to our earlier toy scenario.

\subsection{Model Setup and Symmetry Breaking}

Consider then a model based on the gauge symmetry $SU(N_c)\times SO(10)^4$. The representations of the axion constituents are shown in Table \ref{tableModel2GUT1}, and are all chiral under the weakly-gauged $SO(10)^4$.

\begin{table}[!t]
	\begin{center}
		\begin{tabular}{c||ccccc||ccc}
			 & $SU(N_c)$ & $SO(10)_1$ & $SO(10)_2$ & $SO(10)_3$ & $SO(10)_4$ & $U(1)_{\rm B}$ & $U(1)_{\rm V}$ & $U(1)_{\rm PQ}$ \\
			\hline\hline
			$\psi$ & ${\bf N_c}$ & ${\bf 16}$ & ${\bf 1}$ & ${\bf 1}$ & ${\bf 1}$ & $1$ & $1$ & $1$  \\
			${\widetilde\psi}$ & ${\bf N_c}$ & ${\bf 1}$ & ${\bf 1}$ & ${\bf 16}$ & ${\bf 1}$ & $1$ & $-1$ & $-1$\\
			$\psi^c$ & $\overline{\bf N_c}$ & ${\bf 1}$ & $\overline{\bf 16}$ & ${\bf 1}$ & ${\bf 1}$ & $-1$ & $-1$ & $1$  \\
			${\widetilde\psi}^c$ & $\overline{\bf N_c}$ & ${\bf 1}$ & ${\bf 1}$ & ${\bf 1}$ & $\overline{\bf 16}$ & $-1$ & $1$ & $-1$  
		\end{tabular}
		\caption{Axion constituents and the abelian (continuous) accidental symmetries of the theory.}\label{tableModel2GUT1}
	\end{center}
\end{table}

$SO(10)_1$ is the GUT and the SM fermions (plus the right-handed neutrinos) are as usual embedded in three generations of ${\bf 16}_i$. They are singlets of $SU(N_c)$ and  $SO(10)_{2,3,4}$. As in Section \ref{sec:su5}, other fields charged under $SO(10)_{1}$, but not under $SU(N_c)$ and $SO(10)_{2,3,4}$, can and must be included to obtain a realistic GUT. Specifically, the minimal (non-supersymmetric) set of Higgses belong to the representation $\phi_H\sim{\bf 10}\oplus{\bf 16}\oplus{\bf 45}\in SO(10)_1$. The observed pattern of Yukawa couplings for the SM fermions may be reproduced, as long as \eqref{GUTscale} holds, with a renormalizable coupling to the scalar ${\bf 10}$ and effective dimension-5 operators $\phi^2_H{\bf 16}_i\,{\bf 16}_j/f_{\rm Pl}$. Under these assumptions asymptotic freedom for $SO(10)^4$ implies an upper bound $N_c\leq16$. Because in this new scenario the number of $SU(N_c)$ flavors has grown to $32$, the lower bound on $N_c$ from requiring chiral symmetry breaking becomes $N_c\geq11$. In the end our model lives in the regime 
\ba\label{Nrange}
11\leq N_c\leq16.
\ea
As in the toy model, at $f\ll f_{\rm GUT}$ the $SU(N_c)$ becomes strong and a non-trivial vacuum for the fermion bilinears form:
\ba\label{condensates}
\frac{1}{16}\langle \psi\psi^c\rangle=\frac{1}{16}\langle {\widetilde\psi}{\widetilde\psi}^c\rangle\sim 4\pi f^3\sqrt{\frac{3}{N_c}},
\ea
Switching off the gauge $SO(10)^4$ couplings the particle content of Table \ref{tableModel2GUT1} enjoys a large global $G_{\rm global}\equiv SU(32)_L\times SU(32)_R\times U(1)_{\rm B}$ of independent (non-anomalous) unitary rotations of $(\psi,{\widetilde\psi})$ and $(\psi^c,{\widetilde\psi}^c)$. The weak gauging of $SO(10)^4$ breaks $G_{\rm global}$ explicitly, leaving the global subgroup $SO(10)^4\times U(1)_{\rm B}\times U(1)_{\rm V}\times U(1)_{\rm PQ}$ intact, see Table \ref{tableModel2GUT1}. Below $f_{\rm GUT}$ we can decompose $\psi$ in irreducible representations of $G_{\rm SM,1}$: $\psi=[\psi_{q}\oplus\psi_u\oplus\psi_e]\oplus[\psi_d\oplus\psi_\ell]\oplus[\psi_N]$, with $\psi_N$ a singlet of $G_{\rm SM,1}$ and $SO(10)_{2,3,4}$. The global $SO(10)_1$ is thus further broken into $G_{\rm SM,1}\times U(1)^4$. Without loss of generality the abelian factor $U(1)^4$ can be chosen to act only on the components $\psi_{q,u,e,d,\ell,N}$. It is very simple to identify it. Of the five independent non-anomalous rotations that these fields can be subject to, one is hyper-charge and is already included in $G_{\rm SM,1}$; the remaining four are shown explicitly in Table \ref{tableModel2GUT2}.

\begin{table}[!t]
	\begin{center}
		\begin{tabular}{c||c|ccc}
			& $U(1)_{\rm S, L}$ & $U(1)_{\rm A1}$ & $U(1)_{\rm A2}$ & $U(1)_{\rm A3}$ \\
			\hline\hline
			$\psi_q$ & $-1$ & $1$ & $0$ & $-1$\\
			$\psi_u$ & $-1$ & $1$ & $-1$ & $1$\\
			$\psi_e$ & $-1$ & $-3$ & $-3$ & $1$\\
			$\psi_d$ & $3$ & $1$ & $1$ & $1$\\
			$\psi_\ell$ & $3$ & $-3$ & $0$ & $-1$\\
			$\psi_N$ & $-5$ & $-3$ & $3$ & $1$\\
		\end{tabular}
		\caption{Accidental $U(1)$ global symmetries of the fields of Table \ref{tableModel2GUT1} arising below the GUT scale. }\label{tableModel2GUT2}
	\end{center}
\end{table}

The symmetry breaking pattern induced by the condensates \eqref{condensates} turns out to be
\ba\label{SBP}
&&[G_{\rm SM,1}\times SO(10)_{2}\times SO(10)_3\times SO(10)_4]\\\no
&\times& U(1)_{\rm B}\times U(1)_{\rm V}\times U(1)_{\rm S,L}\times [U(1)^4]_{\rm axial}~~~~~\\\no
&\to& [G_{\rm SM}\times SO(10)_{3+4}]\\\no
&\times& U(1)_{\rm B}\times U(1)_{\rm V}\times U(1)_{\rm S,V},
\ea
where for brevity we introduced the notation 
\ba\label{axial}
[U(1)^4]_{\rm axial}\equiv U(1)_{\rm PQ}\times U(1)_{\rm A1}\times U(1)_{\rm A2}\times U(1)_{\rm A3}
\ea
for the broken axial charges. The condensate $\langle\psi\psi^c\rangle$ breaks $G_{\rm SM,1}\times SO(10)_{2}$ into the vectorial gauge subgroup $G_{\rm SM}$, i.e. the observed SM, whereas $\langle{\widetilde\psi}{\widetilde\psi}^c\rangle$ preserves the diagonal subgroup of $SO(10)_3\times SO(10)_4$. The unbroken abelian group is the linear combination of the original one and $U(1)_{\rm S,R}\subset SO(10)_2$. Indeed, recall that the maximal subgroup of $SO(10)_{2}$ is an $SU(5)\times U(1)_{\rm S,R}$ defined by ${\bf 16}={\bf 10}_{-1}\oplus{\overline{\bf 5}}_{3}\oplus{\bf 1}_{-5}$. Hence, there is a local $U(1)_{\rm S,R}\subset SO(10)_{2}$ that acts on $\psi^c$ exactly as $U(1)_{\rm S,L}$ acts on $\psi^\dagger$, see Table \ref{tableModel2GUT2}. The vectorial combination is left unbroken by the condensates \eqref{condensates}. This is what we denoted by $U(1)_{\rm S,V}$ in \eqref{SBP}.

Without the gauged $SO(10)^4$, the approximate global symmetry breaking pattern $G_{\rm global}\to H_{\rm global}\equiv SU(32)_V\times U(1)_{\rm B}$ produces $1023$ would-be NGBs in the ${\bf 32}\otimes\overline{\bf 32}-{\bf 1}\in SU(32)_V$. We may find their gauge quantum numbers by decomposing the broken generators in irreducible representations of the unbroken gauge group. But it is as instructive to write a more compact expression showing the decomposition under the larger $SO(10)_{1+2}\times SO(10)_{3+4}\subset SU(32)_V$. The result is as follows
\ba\label{NGBs}
{\rm NGBs}&\sim&({\bf 45},{\bf 1})\oplus({\bf 1},{\bf 45})\\\no
&\oplus&({\bf 210},{\bf 1})\oplus({\bf 1},{\bf 210})\\\no
&\oplus&({\bf 16},\overline{\bf 16})\oplus(\overline{\bf 16},{\bf 16})\\\no
&\oplus&({\bf 1},{\bf 1}).
\ea
The NGBs in the first line of \eqref{NGBs} are unphysical and provide the longitudinal components of the broken vectors. The remaining NGBs are all physical. Most of them are charged under $G_{\rm SM}\times SO(10)_{3+4}$ and get a positive mass squared $\sim g^2_{\rm unbroken}f^2/N_c$. There are then four complete singlets that remain massless at the renormalizable level. As we anticipated for the model in Section \ref{sec:su5}, these arise from the breaking of \eqref{axial}. One is obviously the axion in the last line of \eqref{NGBs}, and is associated to the breaking of $U(1)_{\rm PQ}$. The other three are contained in the SM-singlet components of $({\bf 210},{\bf 1})$. They arise from the breaking of the accidental $U(1)_{\rm A1}\times U(1)_{\rm A2}\times U(1)_{\rm A3}$ appearing below the GUT scale. A completely analogous thing happens in the toy model of the previous section, as the reader can verify.

The four singlet NGBs will be denoted by $a,a_1,a_2,a_3$, respectively. They can be described by a 32-dimensional special unitary matrix $\Sigma=e^{2ia_\alpha T_\alpha/f}$ transforming as $\Sigma\to L\Sigma R^*$ under $SU(32)_L\times SU(32)_R$, where the generators, normalized such that ${\rm Tr}[T_\alpha T_\beta]=\delta_{\alpha\beta}/2$, are diagonal and can be readily seen from Table \ref{tableModel2GUT1} and \ref{tableModel2GUT2}. In particular $T_{\rm PQ}=\frac{1}{8}{\rm diag}({\bf 1}_{16\times16}, -{\bf 1}_{16\times16})$ and similarly for $T_{A1,A2,A3}$. The factor of 2 in $\Sigma$ is chosen to conform to the standard definition $\langle0|J_{\rm PQ}^\mu(0)|a(p)\rangle=ifp^\mu$. 

Some of these pseudo-scalars have anomalous couplings to the SM gauge fields proportional to the anomaly coefficients~\footnote{The additional factor of 2 in this expression comes from our conventional factor of 2 in $\Sigma$.} ${\cal A}^{\rm gauge}_\alpha\delta^{AB}=4N_c{\rm Tr}[T_\alpha T_{\rm gauge}^AT_{\rm gauge}^B]$, where $T^A_{\rm gauge}$ are the SM generators, with $A,B$ indices in the adjoint, and the trace runs over all the flavors of the axion constituents. For example, those of $a$ read 
\ba\label{anoma}
-{\delta{\cal L}}_{\rm WZW}&\supset&\frac{g_G^2}{64\pi^2}{\cal A}^G_a\frac{a}{f}\epsilon^{\mu\nu\alpha\beta}G^A_{\mu\nu} G^A_{\alpha\beta}\\\no
&+&\frac{g_W^2}{64\pi^2}{\cal A}^W_a\frac{a}{f}\epsilon^{\mu\nu\alpha\beta}W^i_{\mu\nu} W^i_{\alpha\beta}\\\no
&+&\frac{g_Y^2}{64\pi^2}{\cal A}^Y_a\frac{a}{f}\epsilon^{\mu\nu\alpha\beta}B_{\mu\nu} B_{\alpha\beta}.
\ea
with ${\cal A}^G_a={\cal A}^W_\alpha=N_c$, ${\cal A}^Y_a={5}N_c/{3}$. Similar considerations apply to $a_{1,2,3}$. Despite a unique linear combination of $a,a_{1,2,3}$ can be chosen to couple to the SM gluons, the actual QCD axion is to be identified with $a$. Indeed, as anticipated before and shown explicitly in Section \ref{sec:unstableALP}, the scalars $a_{1,2,3}$ acquire masses above the TeV from dimension-6 interactions. Hence they cannot be relevant to the Strong CP Problem. The only potential impact they can have is to shift $\bar\theta$, an effect which is completely irrelevant to our purposes. On the other hand the QCD axion $a$ remains massless up to a much higher degree, see Section \ref{sec:axion}. Our focus for the rest of the section will therefore be on $a$. Conventionally the scale $f_a$ is defined as
\ba\label{faDEF}
f_a=\frac{f}{{\cal A}^G_a}=\frac{f}{N_c}.
\ea

Before moving on let us note we can verify a posteriori the vectorial gauge group $G_{\rm SM}\times SO(10)_{3+4}$ remains unbroken. All the physical NGBs charged under it have large positive masses squared and, at the renormalizable level, vanishing vacuum expectation values.~\cite{Peskin:1980gc}\cite{Preskill:1980mz} The (misaligning) effect induced by unavoidable higher-dimensional operators is suppressed by at least powers of $f^2/f_{\rm GUT}^2\ll g^2_{\rm unbroken}/16\pi^2$ and hence is much smaller than the (aligning) contributions due to the gauge couplings.

\subsection{UV Contributions to the Axion Potential}
\label{sec:axion}

To relax $\bar\theta$ to zero, and thus solve the Strong CP Problem, we have to make sure that the NGB associated to the breaking of $U(1)_{\rm PQ}$, namely the QCD axion $a$, has a potential dominated by non-perturbative QCD effects. The UV correction $\Delta V$ of eq. \eqref{eq1V} originates from local operators that explicitly break the chiral symmetry $U(1)_{\rm PQ}$. In these models such effects are severely constrained by the local $SO(10)$'s.

The most minimal gauge singlets built out of a ${\bf 16}\in SO(10)$ are either ${\bf 16}\otimes{\bf 16}\otimes{\bf 16}\otimes{\bf 16}$ or ${\bf 16}\otimes\overline{\bf 16}$. With the particle content of Table \ref{tableModel2GUT1} each fermion is charged under a single $SO(10)$, so the combination ${\bf 16}\otimes\overline{\bf 16}$ is automatically invariant under $U(1)_{\rm PQ}$. We are thus left with the former structure, which by construction cannot be $SU(N_c)$ invariant, see \eqref{Nrange}. To form a complete gauge and Lorentz singlet, four powers of $\psi,{\widetilde\psi}$ should be accompanied by at least four powers of $\psi^c,{\widetilde\psi}^c$, and vice-versa. It follows that the most relevant  operators that violate $U(1)_{\rm PQ}$ are of the type
\ba\label{possComb1}
(\psi\psi^c)^4,~~~~~~({\widetilde\psi}{\widetilde\psi}^c)^4,
\ea
and thus belong to the class in eq. \eqref{operGen} with dimension $d=12$. This is a qualitative improvement compared to the model of Section \ref{sec:su5}. We can next employ the general expression \eqref{estimateThetaGen} to estimate the observable vacuum angle in this model. One can check that the operator of type \eqref{possComb1} with the largest $n_{\cal O}q_a$ is $(\widetilde\psi\widetilde\psi^c)^4$, which contributes with $n_{\cal O}=16^4=65\,536$ (!) and $|q_a|=1$, see Table \ref{tableModel2GUT2}. Plugging these values in \eqref{estimateThetaGen}, together with ${\cal A}_a^G=N_c$ from \eqref{faDEF} and $|c_{\rm Pl}|=1$ (the RG effect again leads to an enhancement of a factor of $10^2$), the experimental constraint \eqref{boundTheta} becomes $f_a\lesssim(3-4)\times10^{9}~~{\rm GeV}$, with $N_c$ varying in the range \eqref{Nrange}. Had we not included the large factor of $n_{\cal O}=16^4$, nor the enhancement $({\cal A}_a^G)^{d-1}(3/N_c)^{d/6}\sim N_c^9$, nor the RG enhancement, we would have approached the optimistic $f_a\lesssim10^{11}$ GeV mentioned in the introduction. Unfortunately such enhancements are there. 

Finally we compare the above ``stability" bound on $f_a$ to the lower bound from experimental searches. The low energy phenomenology of our models is in the KSVZ class.~\cite{KSVZ} The effective coupling of the axion to photons, after the mixing with $\pi_0$ is taken into account, is given by $g_{a\gamma\gamma}=\alpha_{\rm em}/(2\pi)[{\cal A}^\gamma_a/{\cal A}^G_a-1.92(4)]/f_a$, where ${\cal A}^\gamma_a\equiv {\cal A}^W_a+{\cal A}^Y_a={\cal A}^G_a(8/3)$. \cite{Srednicki:1985xd} The coupling to neutrons and protons is $g_{ann}=-0.02(3)m_n/f_a$ and $g_{app}=-0.47(3)m_p/f_a$~(see, e.g., \cite{diCortona:2015ldu}). In these scenarios the astrophysics bounds are currently dominated by the supernova constraint on the proton coupling (see e.g. \cite{DiLuzio:2020wdo}) and read $f_a\gtrsim1.3\times10^8~{\rm GeV}$. We conclude that these scenarios have an allowed region of parameter space of one to two orders of magnitude in which the Strong CP Problem is robustly solved.

\subsection{Alternative Realizations}
\label{sec:so10bis}

One can obtain alternative models adding some or removing one of the weakly gauged $SO(10)$ groups. Consider for instance removing one of the gauge factors in Table \ref{tableModel2GUT1}, say $SO(10)_4$. The resulting scenario has a smaller $SU(N_c)\times SO(10)^3$ gauge and the axion constituents are:
\ba\label{PartCont2}
\psi&\sim&({\bf N_c},{\bf 16},{\bf 1},{\bf 1})\\\no
{\widetilde\psi}&\sim&({\bf N_c},{\bf 1},{\bf 1},{\bf 16})\\\no
\psi^c&\sim&(\overline{\bf N_c},{\bf 1},\overline{\bf 16},{\bf 1})\\\no
{\widetilde\psi}^c&\sim&16(\overline{\bf N_c},{\bf 1},{\bf 1},{\bf 1}).
\ea
The physics of the QCD axion is virtually unchanged compared to that of Section \ref{sec:axion}. In particular, also here the leading operators violating $U(1)_{\rm PQ}$ have dimension $12$.

The main difference is that now no additional gauge symmetry survives after chiral symmetry breaking besides the SM, i.e. $SO(10)_3$ is completely broken. Furthermore, as opposed to the model of Section \ref{sec:so10} where only four neutral NGBs arise, here many more are present. These reside in the SM-singlets in $({\bf 16},\overline{\bf 16})\oplus(\overline{\bf 16},{\bf 16})$ and of course in all of the $({\bf 1},{\bf 210})$ of \eqref{NGBs}. Interestingly, among these are potential dark matter candidates, as discussed in the Section \ref{sec:NGBdm}.

\section{Signatures} 
\label{sec:sign}

Our models predict a number of generic signatures: an approximate gauge-coupling unification, heavy unstable ALPs, and dark matter candidates. The first two are essentially the same in the models of Section \ref{sec:su5} and \ref{sec:so10} as well as the one explicitly mentioned Section \ref{sec:so10bis} and will be studied in Sections \ref{sec:unif} and \ref{sec:unstableALP}. Which stable relic can play the role of dark matter depends on the model, though. In the scenario of Section \ref{sec:so10} natural candidates are the lightest glueball of the unbroken $SO(10)_{3+4}$ and/or heavy SM-neutral NGBs; in scenarios of the type of Section \ref{sec:so10bis} dark matter could be in the form of a neutral GeV-scale NGB and/or a heavy hadron again. We will discuss these in Sections \ref{sec:glueballs} and \ref{sec:NGBdm} respectively.

\subsection{Gauge-Coupling Unification at $f_a$?}
\label{sec:unif}

The axion constituents are chiral under the weakly-coupled gauge groups and the SM gauge symmetry observed at low energy, $G_{\rm SM}$, is always in the unbroken vectorial subgroup of the chiral symmetry. An approximate unification of the SM gauge couplings $g_G, g_W, g_{Y'}\equiv\sqrt{5/3}g_{Y}$ at the axion scale follows as a generic prediction. Let us discuss this feature in the context of the $SU(N_c)\times SO(10)^4$ model.

Below the GUT scale the three $G_{\rm SM,1}$ couplings $g_{G,1}$, $g_{W,1}$, and $g_{Y',1}\equiv\sqrt{5/3}\,g_{Y,1}$ start to depart from each other. When we reach $\mu\sim \mu_{\rm gauge}\equiv g_{\rm GUT}f/\sqrt{N_c}=g_{\rm GUT}\sqrt{N_c}f_a$ we find that the observed SM gauge symmetry is in the vectorial subgroup of $G_{\rm SM,1}\times SO(10)_{2}$. Eq. \eqref{SBP} induces a new threshold, which at tree-level \eqref{SBP} reads
\ba\label{threshold-g}
\frac{1}{g_{i}^2}=\frac{1}{g_{i,1}^2}+\frac{1}{g_2^2},~~~~i=G,W,Y',
\ea
with $g_2$ the $SO(10)_2$ and all couplings are renormalized at $\mu\sim \mu_{\rm gauge}$. Now, for generic values $g_2^2\gtrsim g_{i,1}^2$ the threshold correction \eqref{threshold-g} is very mild, if relevant at all. In the complementary regime $g_2^2\lesssim g_{i,1}^2$ (for earlier speculations on this possibility see \cite{Weiner:2001pv})~\footnote{Our gauge coupling unification has also something in common with the string-inspired work \cite{Barbieri:1994jq} and the ``fake GUT" scenario of \cite{Ibe:2019ifm}, though important differences exist.
} the linear combination of vector fields that describes the low energy SM gauge bosons is mostly oriented along the $SO(10)_2$ vectors and the couplings observed at low energy thus have a comparable value $g_i^2= g_2^2[1+{\cal O}(g_2^2/g_{i,1}^2)]$. This results in an {{approximate}} gauge coupling unification at scales $\mu\sim \mu_{\rm gauge}\ll g_{\rm GUT}f_{\rm GUT}$ parametrically smaller than those at which true unification of quarks and leptons actually occurs. Gauge coupling unification is only {\emph{approximate}} because an accurate one would require $g_2^2\ll g_{i,1}^2$, which can only be realized if the couplings $g_{i,1}^2$ are non-perturbative at threshold.

Physics in the approximately unified regime $g_2^2\lesssim g_{i,1}^2$ is quite interesting. The gauge couplings almost unify at the axion threshold, and yet proton decay is not induced because the $SO(10)_2$ gauge bosons do not carry baryon nor lepton charges. One has to probe physics at the GUT scale in order to discover that the SM fermions fill complete $SO(10)_1$ GUT representations. We have thus separated gauge-coupling unification from proton decay. As an extreme limit, one can even push $f_{\rm GUT}$ to the Planck scale, thus suppressing proton decay beyond any foreseeable experimental reach, while preserving an approximate gauge-coupling unification at relatively low scales. \cite{Weiner:2001pv} Looking from bottom-up we see that approximate gauge coupling unification at the axion scale leads to a clear low-energy prediction: there must exist new light particles charged under the SM that make the gauge couplings $g_i$ unify at much lower scales than in usual GUTs. It is not hard to find relics of grand-unified multiplets that can serve our purpose. For example, four Dirac electroweak doublets with hypercharge $\pm1/2$ plus two Dirac singlets with hypercharge $\pm1$, all with masses of order the TeV, would be enough to reach SM gauge coupling unification at $10^{11}$ GeV.

\subsection{Unstable ALPs at the TeV}
\label{sec:unstableALP}

These models feature a unique axial symmetry at the renormalizable level, namely $U(1)_{\rm PQ}$. Yet, below the GUT scale an additional approximate one arises. For definiteness we focus on the model of Section \ref{sec:so10}, where the extra broken generators are those of $U(1)_{\rm A1}\times U(1)_{\rm A2}\times U(1)_{\rm A3}$. Completely analogous considerations apply to the scenario of Sections \ref{sec:su5} and \ref{sec:so10bis}. We are thus interested in the physics of the axion-like particles (ALPs) $a_{\rm ALP}=\left\{a_1,a_2,a_3\right\}$, which we recall are the SM-singlets in $({\bf 210},{\bf 1})$ of \eqref{NGBs}. 

The key observation is that $U(1)_{\rm A1}\times U(1)_{\rm A2}\times U(1)_{\rm A3}$ is violated by the gauge couplings. Indeed, as opposed to $T_{\rm PQ}$, the corresponding generators do not commute with the generators of the heavy vectors, $T_L\in SO(10)_1/G_{\rm SM,1}$. This latter fact explains why that symmetry did not appear above the GUT scale in Table \ref{tableModel2GUT1}. The tree exchange of the heavy vectors therefore leads to dimension-6 interactions of the type $(\psi^\dagger T_L\psi)^2/f_{\rm GUT}^2$ that break $U(1)_{\rm A1}\times U(1)_{\rm A2}\times U(1)_{\rm A3}$ completely. As we emphasized in Section \ref{sec:su5}, this is not sufficient to generate a potential for $a_{\rm ALP}$, though, because such interactions only break $SU(N_f)_L$. The necessary breaking of $SU(N_f)_R$ comes from the gauging of $T_R\in SO(10)_{3,4}$. Combining the dimension-6 operators $(\psi^\dagger T_L\psi)^2/f_{\rm GUT}^2$ with a loop of $SO(10)_{3,4}$ vectors we get a contribution to the potential of $a_{\rm ALP}$, which technically can be parametrized by chiral invariants of the form ${\rm Tr}[T_L^j\Sigma T_R^i\Sigma^\dagger T_L^j\Sigma T_R^i\Sigma^\dagger]$. This translates into a mass matrix (including mass mixing terms) of the order 
\ba\label{a123}
m_{{\rm ALP}}^2&\sim&\frac{g^2}{N_c}\frac{f^4}{f_{\rm GUT}^2}\\\no
&=&(2.5~{\rm TeV})^2\left(\frac{g}{0.6}\right)^2\left(\frac{N_c}{12}\right)^3\left(\frac{f_a}{10^{9}~{\rm GeV}}\right)^4\left(\frac{10^{16}~{\rm GeV}}{f_{\rm GUT}}\right)^4.
\ea
Nothing like this can be written down for $a$, since $U(1)_{\rm PQ}$ is not broken by the gauge interactions.

One can readily see from Table \ref{tableModel2GUT2} that these extra NGBs have anomalous couplings $\propto{\cal A}_{\rm ALP}^{\rm SM}$ to the SM gauge fields ($B_\mu, W_\mu, G_\mu$), see above \eqref{anoma} for an explicit expression of ${\cal A}_{\rm ALP}^{\rm SM}$, and are therefore unstable. Up to unknown mixing angles, their lifetimes are 
\ba
\frac{1}{\tau_{\rm ALP}}=\left(\frac{{\cal A}_{\rm ALP}^{\rm SM}}{{\cal A}_a^G}\right)^2\left(\frac{g_{\rm SM}^2}{64\pi^2f_a}\right)^2\frac{m_{\rm ALP}^3}{\pi}n_{\rm SM}, 
\ea
with $n_{\rm SM}=1,3,8$ for ${\rm SM}=Y, W, G$ a multiplicity factor. Given typical values of the parameters the decay safely takes place at temperatures $T\gtrsim 10^{1-2}$ GeV. Hence, the heavy partners of the QCD axion disappear well before nucleosynthesis and lead to no obvious signatures.

\subsection{Long-lived Relics}
\label{sec:stable}
\label{sec:DWHNBS}


A generic problem of axion models is the production of topological defects via the Kibble-Zurek mechanism. To avoid conflicts with cosmology we assume the highest temperature reached by the Universe, $T_{\rm max}$, never exceeded the critical temperature of the $SU(N_c)$ dynamics, expected to be of order $\Lambda\equiv{4\pi f}/{\sqrt{N_c}}=4\pi f_a\sqrt{N_c}$ by analogy with large N QCD:
\ba\label{reheat}
T_{\rm max}\lesssim 4\times10^9~{\rm GeV}\left(\frac{N_c}{12}\right)^{1/2}\left(\frac{f_a}{10^{9}~{\rm GeV}}\right).
\ea
Under this condition the $SU(N_c)$ dynamics underwent its phase transition during or before inflation (we will see below that also $H_{\rm I}\ll 4\pi f_a$ holds) and topological defects were subsequently inflated away.

Composite axion models also predict a whole tower of heavy hadrons. The lightest baryons have mass $\sim N_c\Lambda$ and are cosmologically stable. Because of \eqref{reheat} their abundance is completely negligible throughout the thermal history of the Universe. Heavy mesons with masses of order $\sim\Lambda$ quickly decay into NGBs as in ordinary QCD. The physics of the NGBs is more model-dependent.

\subsubsection*{Heavy NGBs}

In the scenario of Section \ref{sec:so10} all NGBs besides the QCD axion and the three unstable singlets of Section \ref{sec:unstableALP} are heavy. They decay into SM fermions, gauge bosons, or into fermions and other NGBs.~\footnote{For a heavy fermion singlet $N$ the decay of some of the SM-charged NGBs may take place between BBN and matter domination and destroy the primordial abundance of $^4{\rm He}$ (see e.g. \cite{Kawasaki:2017bqm}). To avoid affecting appreciably physics at BBN we assume at least one of the singlets has $m_N\lesssim10^{11}$ GeV.} The lightest NGB carrying $SO(10)_{3+4}$ charge, we call it $\Pi_s$, is stable. We expect it to be the SM-neutral component of $({\bf 16},\overline{\bf 16})\oplus(\overline{\bf 16},{\bf 16})$ in the third line of \eqref{NGBs}, say $\Pi_s\sim\psi_N\widetilde\psi^c$, since the masses of all its charged partners receive additional positive contributions from the SM gauge loops. A robust way to ensure its abundance does not exceed the density of the observed dark matter, in conflict with the well established hot big-bang theory, is to postulate that the reheating temperature $T_{\rm RH}\leq T_{\rm max}$ is sufficiently smaller than its mass. This constraint roughly reads $T_{\rm RH}<m_{\Pi_s}/30$, where $m_{\Pi_s}^2\sim (g^2_{\rm unbroken}/16\pi^2)C_2\Lambda^2$. Hence we demand
\ba\label{PisDen}
T_{\rm RH}\lesssim10^{8}~{\rm GeV}\left(\frac{g_{\rm unbroken}}{0.5}\right)\left(\frac{N_c}{12}\right)^{1/2}\left(\frac{f_a}{10^{9}~{\rm GeV}}\right),
\ea
with $g_{\rm unbroken}$ the coupling of $SO(10)_{3+4}$ and $C_2=45/8$ the quadratic Casimir of the spinorial of $SO(10)_{3+4}$, the representation carried by $\Pi_s$. With \eqref{PisDen} satisfied, the model of Section \ref{sec:so10} has a negligible abundance of $SU(N_c)$ hadrons. Potential cosmological signatures may come from the $SO(10)_{3+4}$ gauge bosons, as discussed in Section \ref{sec:glueballs}.

The situation is a bit different in the scenario of Section \ref{sec:so10bis}. In that case the NGBs in the $({\bf 210},{\bf 1})$ of \eqref{NGBs} again decay into SM fermions, and those in the third line of \eqref{NGBs} again decay into the lightest of them, i.e. $\Pi_s$. However, $\Pi_s$ and all the states in $({\bf 1},{\bf 210})$ are now massless at tree-level, and couple very weakly to the bath via higher-dimensional operators. Still, $\Pi_s$ may be abundantly produced via the decay of its heavy partners. To avoid this we impose the same condition as \eqref{PisDen}, with $g_{\rm unbroken}$ replaced by the SM couplings. As a consequence of this assumption, also in the scenario of Section \ref{sec:so10bis} the present-day population of heavy hadrons is negligible. The physics of the light, decoupled singlets $\Pi_s$ and $({\bf 1},{\bf 210})$ will be analyzed in Section \ref{sec:NGBdm}. 


\subsubsection*{Axion Density}

Eq. \eqref{reheat} implies the axion abundance is set by vacuum misalignment. Assuming for definiteness that the temperature-dependence of the axion mass is approximately given by $m_a\propto T^{-\delta}$ for temperatures above the critical one, the fraction of dark matter in the form of cold QCD axions turns out to scale as 
\ba\label{axionDM}
\frac{\rho_a}{\rho_{\rm DM}}\sim \theta_{{\rm init},a}^2\left(\frac{f_a}{10^{12} ~{\rm GeV}}\right)^{\frac{\delta+3}{\delta+2}}, 
\ea
where $\theta_{{\rm init},a}$ is the initial misalignment angle. The result \eqref{axionDM} depends uniquely on the initial condition and the (temperature-dependent) QCD potential, and is therefore governed by $f_a=f/N_c$ rather than $f$, see \eqref{faDEF}. The main unknown is $\delta$. Taking as a reference the result of a dilute instanton gas approximation, where $\delta\sim4$, we see that for the relatively low values $f_a\sim10^{8-10}$ GeV considered here the axion constitutes at most a small fraction of the observed dark matter. With $f_a$ fixed, the abundance decreases as $\delta$ decreases.

During inflation the axion field undergoes quantum fluctuations of typical size of the Hubble scale $H_{\rm I}$ in that epoch. The resulting isocurvature perturbations are constrained by cosmological observations. This latter constraint can be expressed as an upper bound on $H_{\rm I}$ \cite{Ade:2015lrj}: $H_{\rm I}\lesssim10^7~{\rm GeV}~(f_a/10^{11}~{\rm GeV})^{(1+\delta)/(4+2\delta)}\sqrt{\rho_{\rm DM}/\rho_a}$, which gets weaker as the QCD axion abundance decreases. 
Combining with \eqref{axionDM} we obtain 
\ba
H_{\rm I}\lesssim\frac{10^8}{\theta_{{\rm init},a}}~{\rm GeV}\left(\frac{10^9~{\rm GeV}}{f_a}\right)^{\frac{1}{2+\delta}}.
\ea
This bound applies to all pre-inflationary QCD axion scenarios. In our case it is a bit relaxed compared to what is often quoted because the axion is a small component of the dark matter.

\subsubsection{Glueball Dark Matter}
\label{sec:glueballs}

The model of Section \ref{sec:so10} predicts an unbroken hidden non-abelian gauge group, i.e. $SO(10)_{3+4}$. Below the scale $\Lambda=4\pi f_a\sqrt{N_c}$ the associated vectors are coupled to ordinary gluons very weakly via higher dimensional operators of the form~\footnote{Dimension-6 operators involving the Higgs doublet are also generated at higher loop order. Despite the lower dimensionality, however, the UV diagrams responsible for generating them are highly convergent. As a consequence their Wilson coefficients are suppressed by the weak scale, i.e. $m_{\rm weak}^2/\Lambda^2$, and their effect turn out to be subleading compared to \eqref{HidGauge} in the interesting region of parameter space where $m_\Phi\gg m_{\rm weak}$. We thank R. Contino for emphasizing this to us.}
\ba\label{HidGauge}
c\frac{N_cg_{3+4}^2g_G^2}{16\pi^2\Lambda^4}~G_{\mu\nu}G^{\mu\nu}~G_{\rho\sigma}'G'^{\rho\sigma},
\ea
and similarly for the electroweak bosons. The interaction is so irrelevant that they never equilibrate with the plasma. The hidden gluons form a gas of self-interacting particles living with its own entropy $s'(T')$ at its own temperature $T'$, which differs from the temperature $T$ of radiation. At the end of inflation the hidden gluons are populated dominantly by the annihilation of SM gluons according to $dn'/dt+3Hn'=\Gamma$, with $H$ the Hubble scale and $\Gamma$ the Boltzmann collision term. (By the assumption \eqref{PisDen} the amount of vectors produced by the annihilation of the hadrons at the scale $\sim\Lambda$ is instead completely negligible.) After being produced, the hidden gas thermalizes right away with entropy $s'(T')=({2\pi^4}/{45\zeta_3})n'(T')$.

The non-renormalizable nature of \eqref{HidGauge} implies $\Gamma\propto T^{12}$, so that the production is mostly active at the highest accessible temperatures $T\sim T_{\rm RH}$. The entropy of the hidden gluons over the radiation entropy may be estimated to be of order 
\ba\label{Yapp}
\xi\equiv\frac{s'}{s}\sim\left.\frac{2\pi^4}{45\zeta_3}\frac{\Gamma}{sH}\right|_{T=T_{\rm RH}}\ll1,
\ea
where we find
\ba\label{Yapp}
{\Gamma}\sim\frac{288}{\pi^5}n_{G}\left(\frac{8\pi^2T^6}{63}\right)^2\left(c\frac{N_cg_{\rm Hid}^2g_{G}^2}{16\pi^2\Lambda^4}\right)^2.
\ea
The quantity $\xi$ then remains constant at all $T\ll T_{\rm RH}$. At later times the $SO(10)_{3+4}$ force confines and the entropy carried by the relativistic hidden gluons gets re-distributed among non-relativistic glueballs. The heavy $SO(10)_{3+4}$ glueballs quickly decay into the lightest, $\Phi$, of mass $m_\Phi$. A simple estimate shows that $\Phi$ has a lifetime longer than the age of the Universe for $m_\Phi\lesssim5\times10^5$ GeV, $c\sim1$, $N_c$ in the range \eqref{Nrange}, and $f_a=10^{8-10}$ GeV. 

The present-day abundance of $\Phi$ can be calculated as follows (see ref. \cite{Carlson:1992fn}). After confinement the new gas of non-relativistic $\Phi$'s remains self-interacting for a while until the number-changing processes go out of equilibrium at $T'_d\lesssim m_\Phi$. We can relate the temperatures of the hidden gas to that of ordinary radiation via $s_\Phi(T'_d)= s'(T'_d)=\xi s(T_d)$. For the range of parameters we are interested in, $T_d$ is always many orders of magnitude above the keV and the energy density $\rho_\Phi$ of the glueballs at decoupling is much smaller than that of radiation. For any $T< T_d$ the non-interacting glueball gas behaves as ordinary cold relics: $\rho_\Phi/s= \xi\rho_\Phi/s'$ stays constant and progressively becomes more and more relevant than radiation. The present-day glueball energy density is of order
\ba\label{DMA}
\frac{\rho_\Phi(T_0)}{s(T_0)}= \xi\frac{\rho_\Phi(T_d')}{s'(T_d')}\sim \xi m_\Phi.
\ea
Using \eqref{Yapp} and saturating the upper bound \eqref{PisDen} the abundance of the stable glueballs reaches a value comparable to the observed dark matter. This conclusion is however extremely sensitive to $T_{\rm RH}$ because of the high temperature-dependence in $\Gamma$. A dedicated study of the reheating process would be needed to clarify if $\Phi$ can be a realistic dark matter candidate in a regime in which the heavy NGBs have a negligible relic density.

\subsubsection{NGB Dark Matter}
\label{sec:NGBdm}

In the scenario of Section \ref{sec:so10bis} there is no hidden gauge symmetry but there is a whole host of light SM-neutral NGBs, as we have seen at the end of Section \ref{sec:DWHNBS}. In particular, $\Pi_s$ and those in the representation $({\bf 1},{\bf 210})$ of \eqref{NGBs} are SM singlets, massless at the renormalizable level, and have no anomalous couplings to the SM gauge fields. Some of these are exactly stable in the renormalizable Lagrangian, namely those that are protected by an ``isospin" symmetry under which the $SO(10)_3$ gauge bosons transform in the adjoint.


A potential is generated by a combination of dimension-6 operators $(({\widetilde\psi}^c)^\dagger\bar\sigma^\mu{\widetilde\psi^c})^2/f_{\rm Pl}^2$, and loops of the massive $SO(10)_3$ gauge bosons. We can recycle \eqref{a123} to get the typical parametric dependence of the masses:
\ba\label{mALP}
m_{{\rm NGB}}^2&=&c_{\rm NGB}\frac{g^2}{N_c}\frac{f^4}{f_{\rm Pl}^2}\\\no
&=&(10~{\rm GeV})^2\left(\frac{c_{\rm NGB}}{1}\right)\left(\frac{g}{0.6}\right)^2\left(\frac{N_c}{12}\right)^3\left(\frac{f_a}{10^{9}~{\rm GeV}}\right)^4.
\ea
A possible decay into SM fermions and gauge bosons requires breaking the ``isospin", triggered by operators suppressed by the Planck scale. Yet, $m_{{\rm NGB}}$ is so small compared to $4\pi f_{\rm Pl}$ that the resulting lifetime is much longer than the age of the Universe. These NGBs are therefore cosmologically stable and may contribute to the dark matter.

Because of \eqref{reheat} and \eqref{PisDen} the stable GeV-scale NGBs have no thermal population and their energy density is set by vacuum misalignment. The standard argument goes as follows. The random initial energy density stayed approximately constant for large $T$. As the temperature dropped below $\sim T_{\rm osc}$, the latter being defined by the condition $3H(T_{\rm osc})=m_{\rm NGB}$, the NGB zero mode started to oscillate and its energy density to decrease as ordinary matter. For typical parameters of our models it turns out that $T_{\rm osc}\gg T_{\rm RH}$. This implies that the NGB field started to oscillate during inflation. Simplifying our calculation assuming a sudden transition at $T_{\rm RH}$, the energy density at later times is given by $\rho_{\rm NGB}=\theta^2_{\rm init}m_{\rm NGB}^2f^2(T/T_{\rm RH})^3/2$. Hence the present-day value is
\ba
\frac{\rho_{\rm NGB}(T_0)}{\rho_{\rm DM}(T_0)}\simeq\left(\frac{\theta_{\rm init}}{10^{-3}}\right)^2\left(\frac{c_{\rm NGB}}{1}\right)\left(\frac{g}{0.6}\right)^2\left(\frac{N_c}{12}\right)^{5}\left(\frac{f_a}{10^{9}~{\rm GeV}}\right)^6\left(\frac{10^{8}~{\rm GeV}}{T_{\rm RH}}\right)^3,
\ea
where the mass was taken from \eqref{mALP}. The bottom line is that the GeV-scale stable NGBs carry too much energy unless their initial displacement is somewhat small. This is reminiscent of the large $f_a$ regime of the ordinary QCD axion, though here $f_a$ is so small that the energy of the QCD axion field is completely negligible. This conclusion is rather conservative, though, because the NGB population could well be depleted significantly during the temperature range $T_{\rm RH}\leq T\leq T_{\rm max}$. In such a case even $\theta_{\rm init}\sim1$ could be acceptable.

\section{Conclusions}

In building composite axion models one knows from the start that the symmetry breaking scale $f_a$ must be very large. Physics at such energies might be more appropriately described in terms of complete multiplets of some GUT. It is therefore only natural to ask whether one can build realistic, asymptotically-free scenarios based on popular grand-unified groups like $SU(5)$ and $SO(10)$. We found that the answer is in the affirmative.

Both $SU(5)$ and $SO(10)$ have chiral representations, the basic ingredient of a working composite axion model. We then looked for scenarios with minimal field content, chiral under $SU(5)$ or $SO(10)$. To avoid a loss of asymptotic freedom the axion constituents were taken to be charged under at most two gauge groups. The existence of an anomalous $U(1)_{\rm PQ}$ of high quality imposed further constraints on the type and number of representations. The simplest model satisfying these requests is based on the ${\bf 10}\oplus\overline{\bf 5}$ chiral multiplet of $SU(5)$. Unfortunately, despite it being quite minimal and elegant, such a scenario allows for UV contributions to the axion potential from operators of dimension 9, which may be too little to obtain a realistic axion model. 

An axion symmetry with higher quality is obtained exploiting the spinorial ${\bf 16}$ of $SO(10)$. In models of this type all $SO(10)$ singlets with $U(1)_{\rm PQ}$ charge contain at least four powers of the axion constituents. As a result the minimal Lagrangian operator that can contribute to the axion potential has dimension 12. We analyzed two models of this class, but many others can be built along these lines.

A rough estimate naively implies that Planck-scale suppressed operators of dimension 12 could be compatible with $f_a\sim10^{11}$ GeV. However, composite axion models tend to have large color and flavor multiplicities --- a must if we want to embed our scenarios in GUTs in which all gauge groups are asymptotically free --- as well as renormalization group effects which go in the direction of enhancing the sensitivity to UV contributions to the potential. Employing naive dimensional analysis, the best tool the effective field theorist has at her or his disposal to estimate the Wilson coefficient of the Planck-scale suppressed operator, we find that our $SO(10)$ models are left with a window of about one to two orders of magnitude $10^8~{\rm GeV}\lesssim f_a\lesssim10^{9-10}$ GeV in which the Strong CP Problem is robustly solved consistently with all experimental data. Our conservative estimate seems to us more justified than the naive one.

Interestingly, the very structure of our scenarios implies a number of very distinctive predictions at low energies. Because of the particular embedding of the SM characterizing these models an approximate gauge coupling unification of (all or just two of) the SM gauge couplings generically occurs at around the axion scale. From a bottom-up perspective this requires new light particles, potentially accessible to colliders, that make the SM gauge couplings unify at $10^{10-11}$ GeV. Above that scale one discovers that the SM gauge fields are mostly the vectors of a grand-unified group, and yet the SM fermions do not fill complete GUT multiplets, so baryon number violating operators do not appear at that threshold. Rather, the actual lepton-quark unification occurs at much larger scales, as usual. 

Another generic prediction of our scenarios is the existence of several light SM-neutral bosonic states beyond the QCD axion. Some of these can be potential dark matter candidates precisely in the regime $f_a\sim10^{8-10}$ GeV expected in our models, i.e. when the QCD axion contributes negligibly. In the models of Section \ref{sec:so10} these dark matter candidates are in the form of long-lived glueballs of an unbroken hidden gauge sector populated at re-heating, whereas in the models of Section \ref{sec:so10bis} they correspond to GeV-scale NGBs with a cold abundance set by vacuum misalignment. In addition we always find unstable ALPs with anomalous couplings to the SM gauge bosons and masses of order a few TeV generated at the GUT scale.

In one way or another, all field theoretic explanations of the axion quality problem look a bit ad hoc. Our $SO(10)$ models are no exception. Yet, taking into account the good quality of their $U(1)_{\rm PQ}$, the natural predisposition to accommodate grand-unified scenarios, and their distinctive phenomenological signatures, we think that our models do not compare too bad with the competition either. Perhaps the ideas developed in this paper will find a more compelling incarnation in the future.


\section*{Acknowledgments}

We would like to thank F. D'Eramo for comments on the cosmological signatures of our models, L. Di Luzio and R. Ziegler for discussions on $SO(10)$ and grand-unification, L. Martucci for conversations on axions in string theory, M. Redi for sharing some of his expertise in composite axion models, I. Shoemaker for suggestions, and especially R. Contino, A. Podo, and F. Revello for comments on the draft and for discussions about their independent forthcoming paper on axions.

\end{document}